\documentclass[12pt,preprint]{emulateapj}
\usepackage{enumerate}

\usepackage{amssymb}

\usepackage{lineno}

\usepackage{rotating}

\usepackage{soul}
\usepackage[usenames,dvipsnames]{color}

\usepackage{url}

\usepackage{hyperref}
\hypersetup{
colorlinks=true,
breaklinks=true,
urlcolor= RedViolet,
citecolor=blue,
linkcolor=blue,
breaklinks=true
}
\usepackage{breakurl}

\usepackage[framemethod=tikz]{mdframed}

\definecolor{vertdc1}{RGB}{20,89,33}
\newcommand{\GG}[1]{} 
 
\newcommand{\REVII}[1]{{{#1}}} 

\slugcomment{}

\shorttitle{Bubbles in Titan's seas: nucleation, growth and RADAR signature}

\shortauthors{Cordier et \textit{al.}}
  
  \def\nature{Nature}\def\natastro{Nat.
  Astron.}\def\natgeo{Nat. Geosci.}\def\AnnderPhys{‎Ann. Phys.
  (Berl.)}\def\icarus{Icarus}\def\pss{Planet. Space Sci.}\def\aap{A\&A}\def\apjl{ApJL}\def\apjs{ApJS}\def\areps{Annu.
  Rev. Earth Planet. Sci.}\def\georl{Geophys. Res. Lett.}\def\jgr{J. Geophys.
  Res.}\def\epsl{Earth Planet. Sci.
  Lett.}\def\nature{Nature}\def\aichej{AIChE J.}
\begin{document}


\title{Bubbles in Titan's seas: nucleation, growth and RADAR signature}

\author{
Daniel~Cordier\altaffilmark{1},
G\'{e}rard~Liger-Belair\altaffilmark{1}
}

\altaffiltext{1}{Groupe de Spectrom\'{e}trie Mol\'{e}culaire et Atmosph\'{e}rique - UMR CNRS 7331
               Campus Moulin de la Housse - BP 1039
               Universit\'{e} de Reims Champagne-Ardenne
               51687 REIMS -- France}

\email{daniel.cordier@univ-reims.fr}
                 
\begin{abstract}
  {In the polar regions of Titan, the main satellite of Saturn, hydrocarbon seas have been discovered by the \textit{Cassini-Huygens}
  mission. RADAR observations have revealed surprising and transient bright areas over Ligeia Mare surface. As suggested by recent
  research, bubbles could explain these strange features. However, the nucleation and growth of such bubbles, together with their
  RADAR reflectivity, have never been investigated. All of these aspects are critical to an actual observation.}
  {We have thus applied the classical nucleation theory to our context, and we developed a specific radiative transfer
  model that is appropriate for bubbles streams in cryogenic liquids.}
  {According to our results, the sea bed appears to be the most plausible place for the generation of bubbles, leading to a 
  signal comparable to observations. This conclusion is supported by thermodynamic arguments and by RADAR properties of a bubbly column. 
  The latter are also valid in the case of bubble plumes, due to gas leaking from the sea floor.}
\end{abstract}


\section{\label{intro}Introduction}

    In 1655, the Dutch astronomer Christiaan Huygens turned his telescope toward Saturn with the intention of studying its rings.
However, to his surprise, besides the rings, he also observed an object that has since been known as the largest 
moon of Saturn: Titan.
More than three centuries after this discovery, Titan still offers surprises. For instance, after the arrival of \textit{Cassini/Huygens}
in the Saturn system, hundreds of lakes and seas of hydrocarbons were detected in Titan's polar regions \citep{stofan_etal_2007}.
One of the northern seas, Ligeia Mare, has shown a strange property: ephemeral RADAR bright areas, nicknamed ``Magic Islands,''
which appear and disappear from one flyby to another \citep{hofgartner_etal_2014,hofgartner_etal_2016}. Several ideas have been 
proposed to explain these transient features. Up to now, only scenarios based on streams of bubbles, due the nitrogen exsolution, 
seem to posses a firm
physical basis \citep{cordier_etal_2017a,malaska_etal_2017}. Indeed, Titan's seas are probably composed of methane and some
ethane, in which atmospheric nitrogen can easily dissolve. The existence of such bubbly plumes is not extravagant, since
bubbles of methane megaplumes are observed in Earth's oceans \citep{leifer_etal_2015,leifer_etal_2017}.
To be efficient RADAR waves reflectors, bubbles must be of a size roughly the same as the RADAR wavelength, 
\textit{i.e.} $2.2$ cm. Here, we focus our purpose on bubbles nucleation and growth, and on bubble plume reflectivity.
This paper is divided into four sections: the first and the second are devoted to the production and evolution 
of nitrogen bubbles, whereas the third concerns the RADAR signature of the bubble streams.
We conclude in the last section.

\section{\label{homonucl}Homogeneous Nucleation of Nitrogen Bubbles}

\begin{figure}[!t]
\begin{center}
\includegraphics[angle=0, width=8 cm]{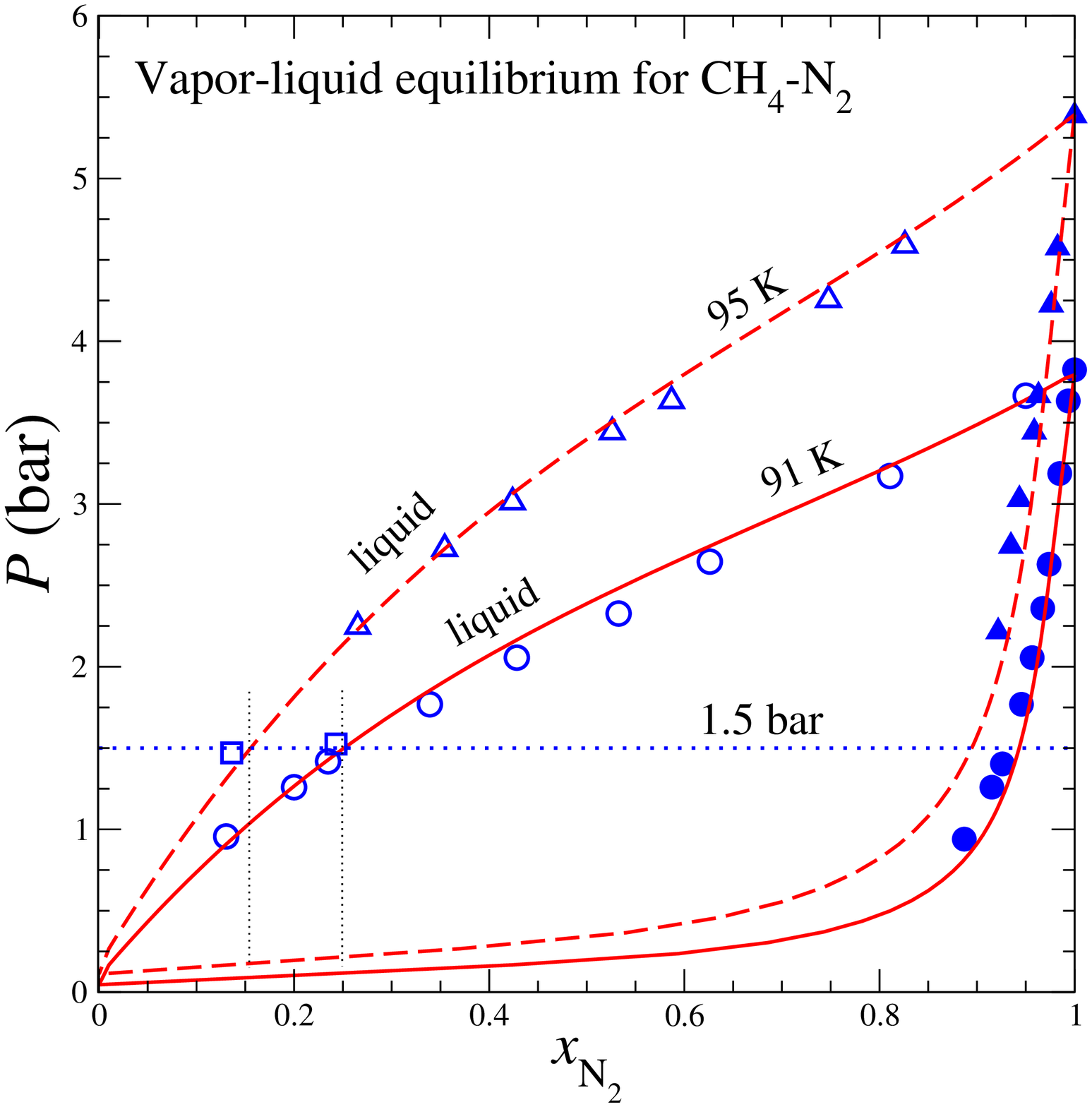}
\caption[]{\label{binaryN2CH4}Comparison between experimental data for the binary system N$_2$--CH$_4$ and our PC-SAFT 
           based model \citep{cordier_etal_2017a}, 
           for two temperatures: $91$ K (circles) and $95$ K (triangles). Laboratory measurements, already used by \cite{tan_etal_2013}, come
           from various sources: \cite{sprow_prausnitz_1966} for $91$ K and \cite{parrish_hiza_1974} for $95$ K (triangles). 
           Squares represent N$_2$ dissolution data from recent work \citep{malaska_etal_2017}, respectively, at $89\pm 0.5$ K and $95\pm 0.5$ K.
           The pressure $P= 1.5$ bar represents the value determined by {\it Huygens} at ground level.}
\end{center}
\end{figure}

    For the sake of simplicity and because this is the most plausible place for a temperature rise to trigger bubbling, we
begin our
reasoning by considering the surface of a Titan's hydrocarbon sea. Then, the relevant
thermodynamic conditions are a temperature within the range of $90-95$ K and a total pressure around $1.5$ bar \citep{cordier_etal_2017a}.
Generally speaking, there are two ways for bubbles to nucleate and grow within a liquid \citep{brennen_1995}. When homogeneous nucleation occurs, the vapor molecules
may come together by collisions, forming embryonic bubbles. Depending on local fluctuations, the vapor deposits around these
embryos and allows some bubbles to grow irreversibly. In the case of heterogeneous nucleation, the vapor molecules add on 
an existing solid substance, foreign in composition to the vapor. In our context, this solid material could be formed by particles in 
suspension into the liquid phase.
    The modern theory of homogeneous nucleation goes back to the early twentieth century 
    \citep{volmer_weber_1926,zeldovich_1943}, 
its results are now well established \citep{brennen_1995}. From this experimental and theoretical corpus, evidence has been provided
to show that an embryo of a bubble has to overcome a ``free energy barrier'' to grow during the nucleation process. This barrier is
well represented by a bubble critical radius $r_{\rm c}$. Bubbles containing gas, with a radius $r_b < r_{\rm c}$, tend to 
redissolve into the liquid phase, whereas embryonic bubbles reaching $r_{\rm c}$ can grow to a much larger size.
The critical radius (in m) is governed \REVII{by Laplace's equation} \citep{brennen_1995}
\begin{equation}
P_{B} - P_{L} = \frac{2 \sigma}{r_{\rm c}}
\end{equation}
%
%
\begin{table}[!t]
\caption{\label{gammas}Surface tensions of the Main Constituents of the Liquid filling the Titan's Seas.
         These data have been provided by the Dortmund Data Bank$^{a}$.}
\begin{center}
\begin{tabular}{l|ccc}
Species                        & N$_2$               & CH$_4$               & C$_2$H$_6$         \\
\hline
$\gamma $(N m$^{-1}$)          & $6 \times 10^{-3}$  & $1.7 \times 10^{-2}$ & $3.15 \times 10^{-2} $ \\
\end{tabular}
\end{center}
\textbf{Notes}. These data have been provided by the dortmund data bank.\\
$^{a}$\url{http://www.ddbst.com}
\end{table}
%
where $P_{B}$ is the pressure (Pa) inside the bubble, $P_{L}$ represents the pressure into the surrounding liquid, and $\sigma$ stands
for the surface tension (N m$^{-1}$). Figure \ref{binaryN2CH4} reports a liquid--vapor equilibria for the system N$_2$--CH$_4$ which 
is relevant, in first approximation, for Ligeia Mare. Two temperatures are considered: $91$ and $95$ K, corresponding to a couple of sets of
measurements. If we restrict our reasoning to the $95$ K case, a liquid under $1.5$ bar could be in equilibrium with
a vapor at a maximum pressure of $\sim 5.4$ bar, composed almost exclusively of nitrogen in that case ($x_{\rm N_2} \sim 1$). From
the difference in pressure $P_{B} - P_{L} \simeq 3.9$ bar, we are able to estimate the corresponding critical radius. According
to surface tension values gathered in Tab.~\ref{gammas}, a cryogenic liquid containing around $20$\% of N$_2$ and $80$\% of
CH$_4$ has a surface tension of $\sigma\sim 1.5 \times 10^{-2}$ N m$^{-1}$. This leads to the critical radius $r_{\rm c} \simeq 10^{-7}$ m.
In principle, other possibilities are conceivable, involving a pressure $P_{B}$ determined between $\sim 5.4$ bar and the liquid pressure of $1.5$ bar.
Clearly, as $P_{B}$ gets closer to $P_{L}$, the critical radius diverges, taking arbitrary large values. However, the net energy required
to form a bubble of radius $r_{\rm c}$ is given by \citep{brennen_1995}
\begin{equation}\label{nuclenergy}
 W= \underbrace{4\pi r_{\rm c}^{2} \sigma}_{\rm(A)} - 
    \underbrace{\frac{4}{3}\pi r_{\rm c}^{3} (P_{B}-P_{L})}_{\rm(B)} = 
    \frac{4}{3}\pi r_{\rm c}^{2} \sigma
\end{equation}
%
%
%
The physical meaning of terms in Eq.~(\ref{nuclenergy}) are the following: (A) represents the energy stored in the surface of the bubble,
while (B) accounts for the work done by the liquid during the bubble inflation. It can be
shown \citep{brennen_1995} that the probability of formation of a microbubble of radius $r_{\rm c}$ is proportional to $\exp -W/k_{\rm B}T$,
with $k_{\rm B}$ the Boltzmann constant. This consideration clearly favors the above mentioned embryonic ($r_{\rm c}\simeq 10^{-7}$ m) bubbles of 
pure nitrogen ($x_{\rm N_2}\simeq 1$), since these small bubbles have a probability of formation much larger than that of bigger bubbles. The theory also provides the homogeneous nucleation rate $J_{\rm nuc}^{\rm hom}$ (m$^{-3}$ s$^{-1}$), 
\textit{i.e.} the mean number of bubbles reaching the critical radius, per unit of volume of liquid, per unit of time \citep{brennen_1995}
\begin{equation}\label{homonucrate}
 J_{\rm nuc}^{\rm hom}= N^{*}_{\rm N_2} \, \left(\frac{2\sigma}{\pi m_{\rm N_2}}\right)^{1/2} \exp - \frac{W}{k_{\rm B}T}
\end{equation}
where $N^{*}_{\rm N_2}$ (m$^{-3}$) is the number of nitrogen molecules, per unit of volume, in the liquid phase, and $m_{\rm N_2}$ (kg) represents the
mass of a single N$_2$ molecule. For the mixture under consideration, we found $N^{*}_{\rm N_2} \sim 3.4 \times 10^{27}$ m$^{-3}$, with an extremely
low nucleation rate 
\begin{equation}\label{lowrateresult}
 \mathrm{log}_{10} J_{\rm nuc}^{\rm hom} \sim -2 \times 10^{5} 
\end{equation}
For a system like Ligeia Mare, which contains roughly $10^{13}$ m$^{3}$ of liquid, the time required for the formation of a single bubble is
much longer than the age of the universe. These estimations unequivocally rule out homogeneous nucleation, as an efficient bubble formation mechanism,
in a Titan's sea. Neither higher pressures nor the presence of ethane changes this conclusion.
At the bottom of a sea, like Ligeia Mare, the pressure is evaluated to be
around $3$ bars \citep{cordier_etal_2017a}, this higher liquid pressure only decreases the difference $P_{B}-P_{L}$ (in Fig. \ref{binaryN2CH4}, $1.5$
bar is replaced by $3.0$ bar) and the nucleation rate is not significantly affected. The presence of some amount of ethane, for instance a mole
fraction of the order of $0.20$--$0.30$ changes only marginally \REVII{the values} of $N^{*}_{\rm N_2}$, while only slightly modifying the surface tension in Eq. (\ref{homonucrate}).

\section{Heterogeneous nucleation and bubbles growth}
%
\begin{figure}[!t]
\begin{center}
\includegraphics[angle=0, width=6 cm]{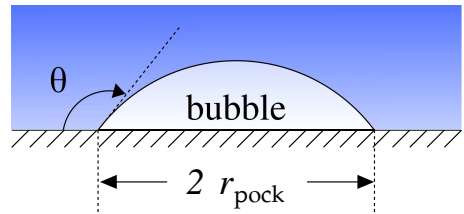}
\caption[]{\label{contactangle}Heterogeneous nucleation on a flat surface, $r_{\rm pock}$ is the typical maximum dimension of the gas pocket. The 
           contact angle at the liquid-vapor-solid intersection is denoted $\theta$ .}
\end{center}
\end{figure}
%
 Alternatively to homogeneous nucleation, heterogeneous nucleation may occur in Titan's seas. It is well known, as a general fact, that 
heterogeneous nucleation is faster than homogeneous nucleation \citep{vehkamaki_2006,sanchezlavega}. 
The presence of a different interface reduces the height of Gibbs free energy barrier.
This is true for all types of phase transition: vapor to liquid, liquid to vapor, liquid to solid, etc. In Titan's seas, the possible
presence of solids may trigger heterogeneous nucleation of nitrogen bubbles. This kind of material could cover the
sea bottom, or could be present under the form of suspended particles. The size of a bubble leaving a solid substrate, under the influence
of buoyancy forces, can be roughly estimated for a contact angle (see Fig.~\ref{contactangle}) around $90^{\rm o}$, value which represents 
the boundary between the low wettability and the high wettability domains. The radius $r_{\rm hetero}$ of the hemispherical vapor nucleus,
leaving its solid horizontal substrate, is given by 
\citep{degennes_etal_2004}
\begin{equation}\label{Rn}
 r_{\rm hetero} \simeq \sqrt{\frac{3\sigma}{\rho g_{\rm Titan}}}
\end{equation}
   If, for example, we consider a $0.8${:}$0.2$ mixture of CH$_4$ and N$_2$, a composition that could be typical of the upper 
layers of liquid,  the surface tension should be $\sigma \simeq 10^{-2}$ N m$^{-1}$ at $90$ K (see Tab.~\ref{gammas}), with a 
density $\rho \simeq 520$ kg m$^{-3}$. These numbers lead to $r_{\rm hetero} \simeq 7 \times 10^{-3}$ m, \textit{i.e.} a diameter of
about $\sim 1.4$ cm. 
If the nucleation occurs at the sea bed, during the rise to the free
surface, bubble will undergo an inflation caused by the pressure drop. Using the law of ideal gases and adopting a pressure of $\sim 3$ bars
at the sea bottom \citep{cordier_etal_2017a}, together with a surface pressure of $1.5$ bar, leads to a radius/diameter enhancement factor of
$2^{1/3} \simeq 1.3$, corresponding to bubbles at the surface with a radius of $1.8$ cm. 
This estimation is more or less comparable to the Cassini RADAR instrument wavelength of $2.2$ cm
Other mechanisms, particularly bubbles coalescence, could also contribute to bubble size evolution, they will be 
discussed in the following. 
It is striking that the video provided by the NASA press release \citep{malaska_etal_2017_NASA_press_release}, 
associated to the Malaska and co-authors article \citep{malaska_etal_2017}, show precisely bubbles leaving a solid 
substrate, which is much larger than bubbles.\\
   Heterogeneous nucleation could also occur on suspended solid particles. To produce cm-sized bubbles, at the moment of solid substrate detachment,
requires solids of similar size. However, such relatively large particles could explain, by themselves, the occurrence of ``Magic Islands'',
without the need for bubbles production, since these preexisting large solids could be good RADAR reflectors \citep{hofgartner_etal_2016}.
In addition, while we know plausible formation processes for bubbles, the presence of solids remains entirely speculative. Therefore, the
formation of cm-sized bubbles, via a purely heterogeneous process is much more plausible at sea bed than anywhere else.
  Nonetheless, the existence of suspended sediments, small enough to be undetectable to the RADAR, cannot be ruled out. Solid particles,
much smaller than the RADAR wavelength may produce embryonic gas pockets, which could grow during their ascent along a column of
liquid. Two distinct growth mechanisms could be at work in such a situation: growth by nitrogen diffusion through bubble surface or
the coalescence of bubbles due to stochastic encounters, within their rising stream. The first possibility requires a liquid
supersaturated in dissolved nitrogen over the entire column, while the second needs a population of bubbles showing a number of
bubbles per unit of volume high enough. We study these two alternative scenarii in the following paragraphs.

\subsection{Bubble Growth by Diffusion}
\label{diff_growth}
   Let us imagine, as suggested by Malaska and co-authors \citep{malaska_etal_2017}, a scenario, according to which a methane-nitrogen Titan's
lake is quickly heated from $\sim 90$ K to $95$ K, \textit{i.e.} fast enough to avoid any degassing. This operation should leave a liquid 
supersaturated in N$_2$. In such a situation, from data plotted in Fig.~\ref{binaryN2CH4}, we conclude that the mole fraction in N$_2$ should be around 
$0.25$ instead of $0.15$, just before the evaporation starts. These mole fractions are respectively equivalent to $7\times 10^{3}$ mol m$^{-3}$ 
and $4.2 \times 10^{3}$ mol m$^{-3}$, leading to a strong supersaturation
of $\Delta c_{\rm N_2}= 2.8 \times 10^{3}$ mol m$^{-3}$. If homogeneous nucleation appears very unlikely, small sediment particles
may generates gas bubbles similar in size to these solid heterogeneities. We have developed a model, that provides the bubbles evolution
during their rise, through layers of liquid hydrocarbons supersaturated in nitrogen. This model, based on the well accepted theory of 
bubbles \citep{clift_etal_1978}, takes into account the bubbles expansion due to pressure drop together with their growth produced
by the diffusion of N$_2$ from the supersaturated liquid to the bubble interior. The details of the model are described in the Appendix.
Our simulations have shown strikingly that the final bubble radius $R_s$, \textit{i.e.} obtained at the surface of the sea, does not depend
on the initial bubble radius $R_0$, but only on the depth $H_0$ at which the embryonic bubble is assumed to start its rise. 
This property is explained by the dependence of rising velocity $U_b$ on bubble radius: $U_b \propto \sqrt{r_b}$. Under this circumstances,
smaller bubbles are the slowest; then, they have more time to let diffusion feeding their interior in nitrogen.
Numerically, we found that a depth of $H_0 \sim 0.5$ m is a minimum to get a radius of $R_s \sim 1.1$ cm at the surface. 
The rise
along such a relatively small height requires only $\tau_{\rm rise} \sim 9$ s. It is clear that, if tiny sediment particles have a
volumic number density high enough, the considered layers of liquid would reach, in a few seconds, the thermodynamic equilibrium with
the atmosphere. Therefore, we have to compare $\tau_{\rm rise}$ with the thermal relaxation time $\tau_{\rm therm}$ of such layers.
In the literature \citep{cordier_etal_2012} we found that $\tau_{\rm therm}$ should be of the order of $2$ Titan's 
days\footnote{One Titan's day corresponds to 15 terrestrial days.}
for $H_0 = 1$ m.
Since $\tau_{\rm therm} \propto H_0^2$, a depth of $H_0 = 0.5 $ m leads to $\tau_{\rm therm} \sim 10^{6}$ s. These numbers suggest that
nitrogen exsolution, by bubbles transport to the surface, should be much faster than thermal relaxation. In such a case, any modest temperature
increase, at the sea surface, would produce an immediate release of nitrogen, under the form of tiny bubbles. As a consequence, liquid layers closest 
to the atmosphere would quickly lose their supersaturation. 
This way, embryonic bubbles, produced in deeper layers, would rise
through non supersaturated zones, a thermodynamic state which inhibits growth by diffusion.
Finally, even if tiny sediment
particles are numerous enough to trigger a quantitative nitrogen dissolution, under the form of small bubbles, the mechanism of growth to
RADAR visible bubbles, should be rapidly blocked by ``de-supersaturation'' of top liquid layers. 
Of course, larger values for $H_0$ make the situation worse.\\
\REVII{A similar desaturation would occur in the case of cosmic rays reaching the surface, even though Titan's dense atmosphere is heavily shielded and the overall cosmic ray flux is low \citep{molina-cuberos_etal_1999}.}
Let us now consider the growth
by bubbles coalescence.

\subsection{Bubble Growth by Coalescence}

   Until this point, we have neglected all possible interaction between bubbles. The features of observed ``Magic Islands'' suggest the
existence of plumes containing a rather large volume density of bubbles. Within a dense population, the probability of the encounters
becomes appreciable. When two bubbles collide, they may coalesce, forming a bigger bubble. This effect substantially enhances the diameter
of bubbles reaching the surface, after having undergone one or several coalescence during the rise. The simplest effect, producing
bubbles collisions, origins in the difference in rise velocities of bubbles of different sizes. The subsequent buoyancy-driven
collision rates $\theta_{ij}^{\rm B}$ (m$^{-3}$ s$^{-1}$) is given by the literature \citep{prince_blanch_1990,friedlander_2000}
\begin{equation}\label{theta_B}
   \theta_{ij}^{\rm B} = n_{i}n_{j} S_{ij} (U_{b,i}-U_{b,j})
\end{equation}
where $n_{i}$ and $n_{j}$ (m$^{-3}$) are the concentration of bubbles of radius $r_{b,i}$ and $r_{b,j}$ (m), and
$S_{ij} =\pi (r_{b,i}+r_{b,j})^2/4$ (m$^2$). Here $U_{b,k}$ is the rise velocity of the particle $k$.
During its ascent, a given bubble $i$ reaches quickly its terminal velocity $U_{b,i}= 2 \sqrt{g_{\rm Titan} r_{b,i}}/3$
(m s$^{-1}$) \citep{clift_etal_1978}. By moving through the liquid, bubbles generate their own, small scale, turbulence, and this expression
of $U_{b,i}$ (also used in the model described in the Appendix) implicitly assumes a turbulent close neighborhood. 
However, as a first approach,
we consider this velocity as an average value and we will take typical radius values in order to get the velocity difference
term in Eq. (\ref{theta_B}), non equal to zero. We have gathered in Tab.~\ref{vitesses_temps} estimations of rising velocities and 
rising timescale $\tau_{100}$ for a $100$ m deep sea, using those radius typical values, \textit{i.e.} $10^{-4}$, $10^{-3}$ and 
$10^{-2}$ m. Since the goal is getting a final bubble with a radius larger than $1$ cm, and
since big bubbles rise faster than small ones (see Tab.~\ref{vitesses_temps}).
faster than small ones, we consider a typical example of a ``test bubble'' of $1$ mm, riding through a population of $0.1$ mm in radius bubbles.
If the differential $\mathrm{d}h$ is the elementary depth variation for our ``$1$-mm bubble'' during
the duration $\mathrm{d}t$, the average number of coalescence events undergone by our ``$1$-mm test bubble'' is
%
%
\begin{table}[!t]
\caption{\label{vitesses_temps}Rising terminal velocities $U_{b,i}$ for three radius values. The Titan's gravity is $g= 1.352$ m s$^{-2}$,
the rising time $\tau_{100}$ is computed for an initial depth of $H_0= 100$ m.}
\begin{center}
\begin{tabular}{l|ccc}
Bubble radius $r_{bi}$ (m)     & $10^{-4}$               & $10^{-3}$               &  $10^{-2}$            \\
\hline
$U_{b,i}$      (m s$^{-1}$)    & $7.8 \times 10^{-3}$    & $2.5 \times 10^{-2}$    & $7.8 \times 10^{-2}$  \\
$\tau_{100}$ (s)               & $1.3 \times 10^{4}$     & $4.1 \times 10^{3}$     & $1.3 \times 10^{3}$   \\
\end{tabular}
\end{center}
\end{table}
%

\begin{equation}\label{dNc}
  \mathrm{d}^2 N_c = \theta^{\rm B}_{ij} \mathrm{d}t \, \mathrm{d}h \, s
\end{equation}
where $s$ represents the cross section of the considered column of liquid, we took $s= 1$ m$^{2}$ for convenience. By integrating 
Eq. (\ref{dNc}) over time and depth, with $\theta^{\rm B}_{ij}$ assumed approximately constant over the entire column, we get 
\begin{equation}\label{Nc}
  N_c \simeq \theta^{\rm B}_{ij} \tau_{100} \, s H_{0}
\end{equation}
If coalescence is the only mechanism at work, a simple calculation, based on the conservation of the total quantity of gas contained 
in bubbles, shows that $N_c \sim 10^{6}$ bubbles, with a radius of $0.1$ mm, are needed to make one final $1$ cm in radius bubble. 
This result can be used to estimate the required order of magnitude of $\theta^{\rm B}_{ij}$, thanks to Eq. (\ref{Nc}), we found
$\theta^{\rm B}_{ij} \sim 1$ coalescences m$^{-3}$ s$^{-1}$, over the column of $H_0 = 100$ m. For one single $1$-mm sized bubble 
(\textit{i.e.} $n_i= 1$) rising along the 
column, we can evaluate the required number density of $0.1$-mm bubbles, needed to get a final centimeter sized bubbles. For that
purpose, Eq. (\ref{theta_B}) is used, together with values available in Tab.~\ref{vitesses_temps}, to finally obtain 
$n_j \sim 10^{8}$ bubbles per m$^3$ along the entire column (with $r_{b,j} \sim 10^{-4}$ m). 
Polydisperse bubbles populations may be simply generated by sea floor composition heterogeneities, or caused by stochastic fluctuations 
in bubble/substrate uncoupling. Clearly, buoyancy-driven bubbles coalescence appears to be an efficient mechanism which could produce cm-sized
bubbles in bubbles trajectory ends, within the sea top-layers.

This concentration represents $100$ bubbles per cm$^3$,
\textit{i.e.} a total volume of gas of $4.2 \times 10^{-4}$ cm$^{3}$, per cm$^{3}$ of liquid. Number which appears reasonable, since it corresponds
only to a small fraction of the volume of liquid. In this scenario, bubbles are formed at the sea bed, with a non-uniform distribution in size.
The first bundles of bubbles, leaving the depths of the sea, settle all the sea levels. The following generations of bubbles pass through this bubbly medium.
As big bubbles are faster than small ones, similarly to our example, big bubbles (\textit{e.g.} with $r_{b,i} \sim 10^{-3}$ m) aggregate
small ones (\textit{e.g.} with $r_{bi} \sim 10^{-4}$ m).\\

  Collisions induced only by different rising velocities assume a gentle turbulent field, mainly localized in the immediate vicinity of bubbles.
In massive bubbles streams, a strong turbulent field may appear. Under such regime, the turbulent collision rate $\theta^{T}_{ij}$
(see Eq. \ref{theta_T}) no longer depends on differences of individual velocities. Instead, it can be estimated with \citep{prince_blanch_1990}
\begin{equation}\label{theta_T}
   \theta_{ij}^{\rm T} = 0.089 \pi n_{i}n_{j} (d_{bi}+d_{bj})^2 \epsilon^{1/3} (d_{bi}^{2/3}+d_{bj}^{2/3})^{1/2}
\end{equation}
where $d_b$ is the bubble diameter and $\epsilon$ is the energy dissipation per unit of mass and unit of time (J kg$^{-1}$ s$^{-1}$).
Compared to buoyancy-driven collision, in that case, even bubbles of the same size can coalesce. The factor $\epsilon$
can be estimated using $k_{d}= \epsilon^{1/4}/2 \nu^{3/4}$ where $k_d$ (m$^{-1}$) is the wave number of turbulent eddies and 
$\nu$ is the liquid kinetic viscosity \citep{batchelor_1953}. Eddies most affecting bubbles have wave numbers roughly similar
to $1/r_b$. Density and viscosity of liquid methane, together with an assumed radius of $10^{-3}$ m, yields to $\epsilon \sim 0.4$ J kg$^{-1}$ s$^{-1}$.
Assuming a population of mm sized bubbles, basic computations show that we need a number of $N_c= 10^{3}$ of such bubbles 
to build up a cm size final bubble by successive coalescences. For a sea depth of $H_0= 100$ m, corresponding to $\tau_{100}\sim 10^{3}$ s
for a mm size particle, we found the required turbulent collision rate to be of the order of $\theta_{ij}^{\rm T} \sim 10^{-2}$ m$^{-3}$ s$^{-1}$.
We, then, can derive the minimum bubbles density $n_j \sim 10^{5}$ m$^{-3}$. This represents one single mm size bubble for $10$ cm$^{3}$, which
is a pretty modest concentration. Here, the volume of gas is also, incidentally, equal to $4.2 \times 10^{-4}$ cm$^{3}$ per cm$^{3}$ of liquid.\\

  As we can see, both coalescence mechanisms, are able to produce bubbles big enough to be detectable at Ligeia Mare surface. This
conclusion is true if the number of bubbles initially produced is sufficiently large and if their start their journey to the surface from
a depth of the order of $\sim 100$ m. These two coalescence processes may be at work in nature, depending on the size distribution and 
volumic density by number of bubble populations, initially nucleated at the seabed. 
We emphasize that turbulence-driven collision rate could dominate if gases are injected in the liquid through hypothetical sea bottom vents. 
In that case, high $\epsilon$ values could be reached, causing large collision rate. Finally, we stress that break-up radius 
\citep{clift_etal_1978,cordier_etal_2017a} $r_{\rm bk} \simeq 4 \sqrt{\sigma/(\rho \, g)}$
%
%
  cannot be overcome by any mechanism. This remains an absolute upper limit, of the order of 
$\sim 2.3$ cm ($\sim 4.6$ cm in diameter) \citep{cordier_etal_2017a}, for bubble size.
%
\section{Bubbles RADAR Signature}

     Throughout the discussion, the criterion used to decide whether or not bubbles could be RADAR detectable is based on their size.
Objects possessing a diameter comparable to the wavelength (\textit{i.e.} $2.16$ cm) have been considered to have a measurable
effect. This approach is relevant in first approximation, but it is not---by essence---not quantitative, and it neglects effects like
multiple scattering, which may be important in the context. Previous works \citep{hofgartner_etal_2016} have estimated the possible
single scattering albedo of a population of relatively small bubbles ($r_b \sim 10^{-3}$ m), \textit{i.e.} using the Rayleigh
scattering theory \citep{bohren_huffman_2004}. For larger reflectors, \textit{i.e.} with sizes comparable to the wavelength, the Mie 
scattering theory is required \citep{mie_1908}.
%
\begin{figure}[t]
\begin{center}
\includegraphics[angle=0, width=7 cm]{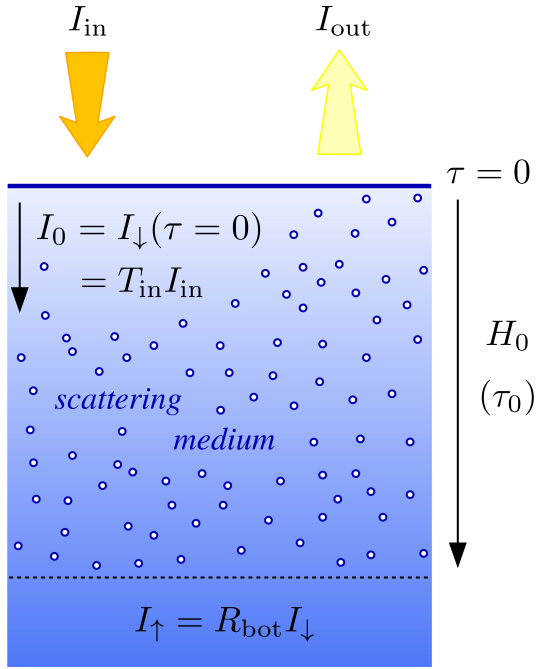}
\caption[]{\label{colBubbles}Sketch of the liquid column containing bubbles. $I_{\rm in}$ is the incident RADAR photons flux, while
            $I_{\rm out}$ is the signal returning back to the emitter. The total height of the column is denoted $H_{0}$ corresponding
            to the optical depth $\tau_{0}$. The bottom of the column, filled by bubbles, does not necessarily correspond to the sea bed.}
\end{center}
\end{figure}
%
  We then built a model in which a column of liquid is filled by bubbles, with a total height denoted $H_0$, corresponding to the ``optical'' 
depth $\tau_{0}$ (see Fig.~\ref{colBubbles}). The geometry is simplified: the flux of energy $I_{\rm in}$ coming from the spacecraft 
arrives at the sea-atmosphere interface with a normal incidence. This approximation is perfectly relevant in our case, since during
T92 and T104 observations, the incident angles were respectively $6.^{\rm o}0$ and $11.^{\rm o}5$ \citep{hofgartner_etal_2016}.
The effects of the polarization, and the absorption, are neglected as suggested by previous works 
\citep{hofgartner_etal_2016}. In that frame, using a two-stream radiative transfer model \citep{bohren_huffman_2004}, accounting for 
multiscattering by principle, the energy fluxes, through the liquid, in downward and upward direction are, respectively,
\begin{equation}
I_{\downarrow}= D + C (1 - \tau^*)
\end{equation}
\begin{equation}
I_{\uparrow}= D - C (1 + \tau^*)
\end{equation}
where $\tau^*$ is the optical depth corrected by the asymmetry factor $g_b$ of bubbles: $\tau^*= (1-g_b) \tau$. The asymmetry factor is computed
in the frame of the Mie's theory. The coefficients $D$ and $C$ are given as a functions of $I_0= I_{\downarrow, \tau=0}$, and $\tau^*_0$ 
the total optical depth of the column (see Fig.~\ref{colBubbles}), we have
\begin{equation}
C = \frac{(1-R_{\rm bot}) I_0}{2 + (1-R_{\rm bot}) \tau^*_0}
\end{equation}
\begin{equation}
D = \frac{ 2 + (1-R_{\rm bot}) (\tau^*_0 - 1)}{ 2 + (1-R_{\rm bot}) \tau^*_0} I_0
\end{equation}
where $R_{\rm bot}$ represents the reflectance at the bottom of the column, or equivalently at the sea floor.
The uncorrected total optical depth $\tau_0$, of the column of the liquid, is provided by
\begin{equation}
    \tau_0= \int_{0}^{H_0} \beta \, \mathrm{d}z
\end{equation}
in which $1/\beta$ represents the radar photon's mean-free path; $\beta$ is a function of the number density $n_b$ (bubbles m$^{-3}$) and of
the bubbles Mie's cross section $\sigma_{\rm Mie}$: $\beta= n_b \, \sigma_{\rm Mie}$. The flux leaving the sea and returning to the RADAR is
$I_{\rm out}= (1-R_{\rm in}) I_{\uparrow, \tau=0}$, here,
the reflectance $R_{\rm in}$ of the interface sea atmosphere, is assumed to take into account the effect of the rugosity
\citep{grima_etal_2017},
which is usually, except in the occurrence of a ``Magic Island'' event, measured to be very small
\citep{wye_etal_2009,zebker_etal_2014,grima_etal_2017,stephan_etal_2010,barnes_etal_2011b}.\\
       In order to quantify the RADAR signature of bubbles, we compare the reflected flux with and without the presence of bubbles. For that
purpose, we introduce the quantity $R_{\rm bubb}=I_{\rm out}/I_{\rm in}$ which has to be compared to the `` clear sea,'' \textit{i.e.}
without bubbles, global reflectance given by $R_{\rm cs}= R_{\rm bot} T_{\rm in}^{2}+ R_{\rm in}$, where $T_{\rm in}= 1 - R_{\rm in}$.
For that purpose, we denote Bubbles Radar Signal Amplification (BRSA)  as the ratio $R_{\rm bubb}/R_{\rm cs}$.\\

\begin{figure*}[t]
\begin{center}
\includegraphics[angle=0, width=17 cm]{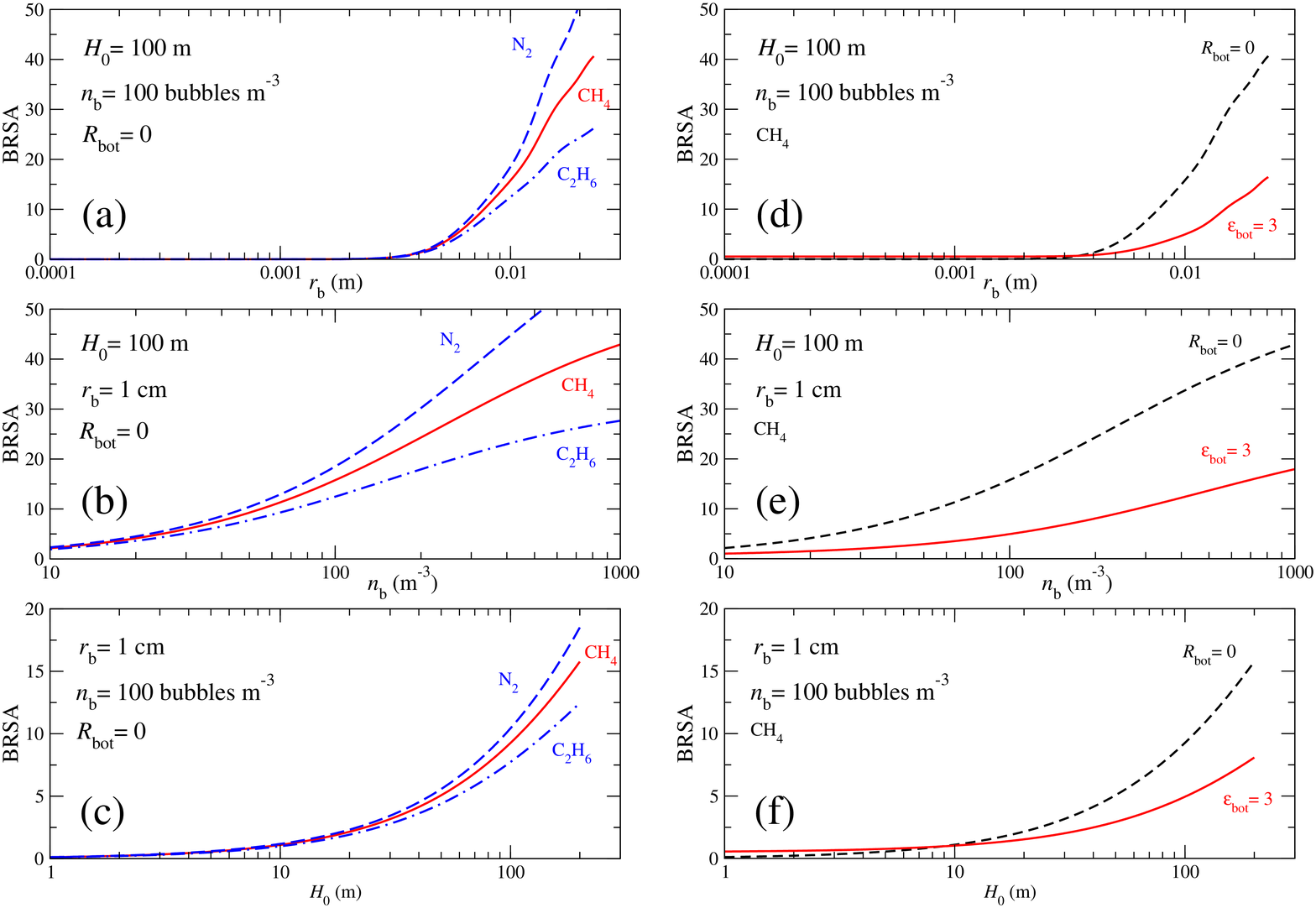}
\caption[]{\label{influ_param_RADAR}Influence of bubble stream parameters on RADAR Signal Amplification (BRSA).
           (a) Influence bubble radius $r_b$, the simulation is stopped at the bubbles break-up radius $r_{\rm breakup}\sim 2.3$ cm
           \citep{cordier_etal_2017a}. While the solid red curve has been computed using the methane permittivity $\epsilon_r= 1.72$ 
           \citep{mitchell_etal_2015}, dashed and dotted-dashed blue lines correspond, respectively, to liquid nitrogen 
           \citep[$\epsilon_r= 1.55$;][]{hosking_etal_1993} and ethane \citep[$\epsilon_r= 2.00$;][]{mitchell_etal_2015}.
           (b) Influence of volume density $n_b$ of bubbles. (c) Influence of the
           total height of the column $H_0$ (see Fig.~\ref{colBubbles}), the considered range of $H_0$ has been limited to
           $0-200$ m, since the bathymetry of Ligeia Mare shows a maximum depth around $200$ m \citep{hayes_2016}. In panels (d), (e)
           and (f), we report computations including a non-zero reflectance of the sea floor; its permittivity is $\epsilon_{\rm bot}= 3$,
           which is probably a very high value (corresponding to $R_{\rm bot} \simeq 2$\%. In these panels, the methane $R_{\rm bot}=0$
           curve is recalled for comparison.}
\end{center}
\end{figure*}

   As a first approach, we have chosen to neglect the upward flux of RADAR photons at the bottom of the column: $I_{\uparrow}(\tau_0)= 0$.
Below the bubbly column, the microwave photons are considered to be lost. In other words, the reflectance at 
the bottom of the column is taken equal to zero: $R_{\rm bot}= 0$. Taking the methane
permittivity \citep{mitchell_etal_2015} $\epsilon_r({\rm\tiny CH_4})= 1.72$ as a reference, we have explored the influences of the bubble
radius $r_b$, of the number of bubbles per unit of volume $n_b$ and of the column height $H_0$, results are gathered in Fig.~\ref{influ_param_RADAR}.
Not surprisingly, large bubble radii favor a strong backscattering (Fig.~\ref{influ_param_RADAR} a). Similar effects are found
for the influence of the number of bubbles per unit of volume $n_b$ (Fig.~\ref{influ_param_RADAR} b) and the total height $H_0$ of the bubbly 
column (Fig.~\ref{influ_param_RADAR} c).
%
%
   The dielectric permittivity of the liquid also has its influence. Taking the permittivity of pure liquid nitrogen:
$\epsilon_r({\rm N_2})= 1.55$ \citep{hosking_etal_1993}, we found a BRSA higher than values obtained with CH$_4$ permittivity
(see Fig.~\ref{influ_param_RADAR}, panels (a), (b), (c)). In contrast, a simulation with liquid ethane permittivity, 
$\epsilon_r({\rm C_2H_6})= 2.00$ \citep{mitchell_etal_2015}, yields to a reduction of the BRSA. Perhaps surprisingly, a low 
liquid permittivity favors the bubble stream RADAR reflection. The chemical composition of Titan is still not firmly known, but
we emphasize that, accidentally, the mean value of nitrogen and ethane respective permittivities is very close to the methane individual value.
Consequently, a sea with a composition in N$_2:$CH$_4:$C$_2$H$_6$ around $0.20:0.40:0.20$ will show a permittivity close 
to the pure liquid methane value $\epsilon_r({\rm CH_4})= 1.72$ \citep{mitchell_etal_2015}.
    In Ref.~\cite{hofgartner_etal_2016}, the Normalized Radar Cross Section (NRCS) along the flyby tracks is
reported in Figures 4 and 5. In these figures, the NRCS ``peaks'' corresponding to T92 and T104 transient feature events offer the opportunity
to estimate the ratio of the quantity of radar photons backscattered with a the presence of a ``Magic Island'' and without such a structure.
The height of NRCS ``peaks,'' measured to be between $\sim 6$ and $\sim 9.5$ in dB, leads to ratios ranging between $\sim 6$ and $\sim 10$. 
This means that radar reflectors present at Ligeia Mare, during ``Magic Island'' episodes, enhance the local reflectivity by a factor in the 
interval $6-10$. Panel (c) in Fig. 6 of the same reference, gives another opportunity to evaluate the ``reflectivity enhancement'' during
Ligeia Mare overbrightness events. A quick comparison of NRCS predicted by the sea floor model plotted in this figure and actual measurements
performed during T92 and T104, leads to energy ratios magnified by a factor of $\sim 10-16$. If we keep a factor around $\sim 10$,
which corresponds to what we call BRSA, the Ligeia Mare ``Magic Islands'' can be easily explained by a column of $H_0 \sim 100$ m, containing
around $100$ centimetric bubbles per cubic meters, this, if sea floor reflectance can be neglected.\\
  Unfortunately, the hypothesis of the seabed zero-reflectivity is an oversimplification. Actually, the sea floor partly re-emits the
incident RADAR beam energy. This property has been utilized to derived Ligeia Mare bathymetry \citep{hayes_2016}. For that
purpose, two distinct echoes in altimetry tracks \citep{hayes_2016} have been detected \citep{hayes_2016}, one caused by the surface and the second
produced by energy backscattered by the sea bottom. Thus, we have compared published values of NRCS \citep{hayes_2016} of these echoes; we found
a difference in dB around $30$, which leads to a ratio in energy of about $\sim 10^{3}$. The flux coming from the deepest part of the 
sea is obviously the weakest, suggesting a quite low reflectance of the sea bottom. Using RADAR observation, and their models, 
Hofgartner and co-authors \citep{hofgartner_etal_2016} propose a sea floor dielectric constant around $\epsilon_{r,{\rm seafloor}}= 1.99$, but
the actual value is not well constrained since the real nature of the seabed is not known. Titan belongs to the so-called ``icy moons;'' therefore,
water ice is recognized to be a major component of Titan's geological layers \citep{balland_etal_2014}. If we assume a sea floor composed by
pure water ice, its microwaves permittivity should be around $\epsilon_{r,{\rm ice}}= 3$ \citep{bradford_etal_2009}. The actual value depends
on the porosity of the ice and on the nature of the material mixed within it. 
Adopting $\epsilon_{r,{\rm ice}}= 3$ for the seabed, which
has to be understood as a high value \citep{legall_etal_2016}, we computed the corresponding BRSAs. They are compared to their counterparts computed with a bottom 
zero-reflectivity; results are plotted in panels (d), (e), and (f) of Fig.~\ref{influ_param_RADAR}. These simulations demonstrate that a non-zero bottom
reflectivity ($R_{\rm bot} \ne 0$) damps the BRSAs, \textit{i.e.} the ratios $R_{\rm bubb}/R_{\rm cs}$. This behavior is caused by the addition
of the term $R_{\rm bot}T^{2}_{\rm in}$ in the expression of $R_{\rm cs}$.
%
\begin{figure}[t]
\begin{center}
\includegraphics[angle=0, width=9 cm]{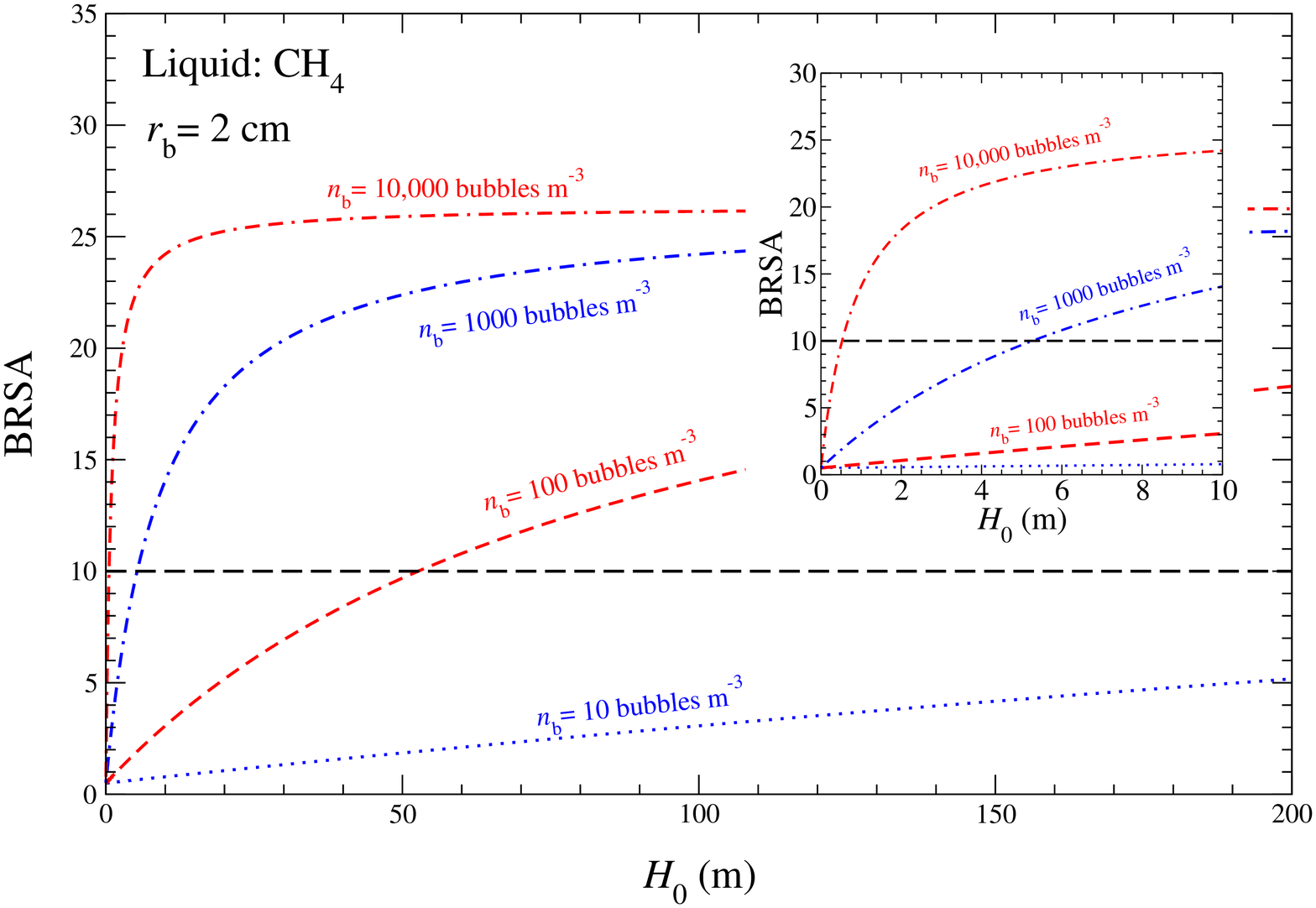}
\caption[]{\label{BRSA}The Bubbles Radar Signal Amplification (BRSA) for a column of liquid methane, harboring bubbles with a radius
           of $2$ cm, as a function of the total height $H_0$ of the column. Several volume densities of bubbles are considered:
           $n_b= 10$, $100$, $1000$, and $10{,}000$ bubbles m$^{-3}$. The explanation of Ligeia Mare ``Magic Island'' requires
           BRSA$\sim 10$. The panel inserted on the right-hand side is a magnification of the main figure in the region of the
           origin.}
\end{center}
\end{figure}
%
  Nonetheless, as we can see in Fig.~\ref{influ_param_RADAR}, even with a relatively large value for $R_{\rm bot}$ ($\epsilon_{r,{\rm seafloor}}= 3$
corresponds to $R_{\rm bot} \sim 2$\%), reasonable combinations of $r_b$, $n_b$ and $H_0$ can be found, with a resulting BRSA around $\sim 10$, 
a value that explains the observed Ligeia Mare ``Magic Island.'' For instance, a column of $H_0= 100$ m, containing $n_b= 100$ bubbles m$^{-3}$ with
$r_b= 2$ cm has a BRSA of $\sim 14$. If we adopt a bubble radius close to the maximum value allowed by bubble physics, \textit{i.e.}
$r_b= 2$ cm which is approximately the break-up radius, we can search for the minimum height $H_0$ required to get a BRSA around $\sim 10$. This
is done in Fig.~\ref{BRSA}, in which several values for $n_b$ are assumed. Since with $r_b= 2$ cm, one cannot include more than
$\sim 10,000$ bubbles within one cubic meter, $n_b= 10{,}000$ bubbles m$^{-3}$ represents a geometrical maximum. As we can see, even with
$r_b= 2$ cm and $n_b= 10{,}000$ bubbles m$^{-3}$, we need $H_0 \sim 0.5$ m to reach BRSA$\sim 10$. According to the discussion conducted
in Sect.~\ref{diff_growth}, it appears impossible to form centimetric bubbles due to an heating starting at the sea surface. One more
time, a scenario based on a bubbles production in the depth of Ligeia Mare looks more plausible than a pure surface phenomenon. Indeed,
Fig.~\ref{BRSA} tells us that a few tens of meters, with a relatively modest number of bubbles per cubic meters, produced the 
required value for the Bubbles Radar Signal Amplification.

\section{Conclusion}

     In this work, we have demonstrated that the homogeneous nucleation of small bubbles of N$_2$ is impossible under the conditions of the
Titan surface. Heterogeneous nucleation, \textit{i.e.} involving a solid substrate, is much more easier. Such substrates could be found
at the seabed or under the form of small sediment particles suspended in the liquid. However, in that case, a growth mechanism has to be
at work to obtain bubbles large enough to be efficient RADAR reflectors. While the growth by diffusion in nitrogen supersaturated layers
appears to be very difficult, if not impossible; the growth by coalescence, along a bubbly column has been found to be a powerful process
to get large bubbles. In this case, such a column must have a height that is more or less comparable to Ligeia Mare depth. We also developed
a model of reflection of the RADAR wave by a stream of bubbles in a Titan's sea. This approach also favors streams of bubbles with a 
vertical extension of several tens of meters.\\
  In short, to explain the ``Magic Islands,'' one scenario, based on bubbles, has the best plausibility if it implies that bubbles are 
released or formed in the depths of the sea.



\begin{thebibliography}{}
\expandafter\ifx\csname natexlab\endcsname\relax\def\natexlab#1{#1}\fi

\bibitem[{{Baland} {et~al.}(2014){Baland}, {Tobie}, {Lef{\`e}vre}, \& {Van
  Hoolst}}]{balland_etal_2014}
{Baland}, R.-M., {Tobie}, G., {Lef{\`e}vre}, A., \& {Van Hoolst}, T. 2014,
  \icarus, 237, 29

\bibitem[{{Barnes} {et~al.}(2011){Barnes}, {Soderblom}, {Brown}, {Soderblom},
  {Stephan}, {Jaumann}, {Mou{\'e}lic}, {Rodriguez}, {Sotin}, {Buratti},
  {Baines}, {Clark}, \& {Nicholson}}]{barnes_etal_2011b}
{Barnes}, J.~W., {Soderblom}, J.~M., {Brown}, R.~H., {et~al.} 2011, \icarus,
  211, 722

\bibitem[{{Batchelor}(1953)}]{batchelor_1953}
{Batchelor}, G.~K. 1953, {The Theory of Homogeneous Turbulence} (Cambridge:
  Cambridge University Press)

\bibitem[{{Bohren} \& {Huffman}(2014)}]{bohren_huffman_2004}
{Bohren}, C.~F., \& {Huffman}, D.~R. 2014, {Absorption and Scattering of Light
  by Small Particles}, 2nd edn. (Wiley-VCH)

\bibitem[{{Bradford} {et~al.}(2009){Bradford}, {Harper}, \&
  {Brown}}]{bradford_etal_2009}
{Bradford}, J.~H., {Harper}, J.~T., \& {Brown}, J. 2009, Water Resour. Res.,
  45, W08403

\bibitem[{{Brennen}(1995)}]{brennen_1995}
{Brennen}, C.~E. 1995, {Cavitation and Bubble Dynamics} (Oxford University
  Press)

\bibitem[{{Clift} {et~al.}(1978){Clift}, {Grace}, \& {Weber}}]{clift_etal_1978}
{Clift}, R., {Grace}, J.~R., \& {Weber}, M.~E. 1978, {Bubbles, Drops and
  particles} (New York, San Fransisco, London: Academic Press)

\bibitem[{{Cordier} {et~al.}(2017){Cordier}, {Garc{\'{\i}}a-S{\'a}nchez},
  {Justo-Garc{\'{\i}}a}, \& {Liger-Belair}}]{cordier_etal_2017a}
{Cordier}, D., {Garc{\'{\i}}a-S{\'a}nchez}, F., {Justo-Garc{\'{\i}}a}, D.~N.,
  \& {Liger-Belair}, G. 2017, \natastro, 1, 0102

\bibitem[{{Cordier} {et~al.}(2012){Cordier}, {Mousis}, {Lunine}, {Lebonnois},
  {Rannou}, {Lavvas}, {Lobo}, \& {Ferreira}}]{cordier_etal_2012}
{Cordier}, D., {Mousis}, O., {Lunine}, J.~I., {et~al.} 2012, \pss, 61, 99

\bibitem[{{de Gennes} {et~al.}(2004){de Gennes}, {Brochard-Wyart}, \&
  {Qu\'{e}r\'{e}}}]{degennes_etal_2004}
{de Gennes}, P.-G., {Brochard-Wyart}, F., \& {Qu\'{e}r\'{e}}, D. 2004,
  {Capillarity and Wetting Phenomena: Drops, Bubbles, Pearls, Waves} (New York:
  Springer), doi:10.1007/978-0-387-21656-0

\bibitem[{{Forster}(1963)}]{forster_1963}
{Forster}, S. 1963, Cryogenics, 3, 176

\bibitem[{{Friedlander}(2000)}]{friedlander_2000}
{Friedlander}, S.~K. 2000, {Smoke, Dust and Haze: fundamentals of aerosol
  dynamics}, 2nd edn. (Oxford: Oxford University Press)

\bibitem[{{Grima} {et~al.}(2017){Grima}, {Mastrogiuseppe}, {Hayes}, {Wall},
  {Lorenz}, {Hofgartner}, {Stiles}, {Elachi}, \& {Cassini Radar
  Team}}]{grima_etal_2017}
{Grima}, C., {Mastrogiuseppe}, M., {Hayes}, A.~G., {et~al.} 2017, \epsl, 474,
  20

\bibitem[{{Hayes}(2016)}]{hayes_2016}
{Hayes}, A.~G. 2016, \areps, 44, 57

\bibitem[{{Hellemans} {et~al.}(1970){Hellemans}, {Zink}, \& {Van
  Paemel}}]{hellemans_etal_1970}
{Hellemans}, J., {Zink}, H., \& {Van Paemel}, O. 1970, Physica, 46, 395

\bibitem[{{Hofgartner} {et~al.}(2014){Hofgartner}, {Hayes}, {Lunine}, {Zebker},
  {Stiles}, {Sotin}, {Barnes}, {Turtle}, {Baines}, {Brown}, {Buratti},
  {Clarck}, {Encrenaz}, {Kirk}, {Le Gall}, {Lopes}, {Lorenz}, {Malaska},
  {Mitchell}, {Nicholson}, {Radebaugh}, {Wall}, \&
  {Wood}}]{hofgartner_etal_2014}
{Hofgartner}, J.~D., {Hayes}, A.~G., {Lunine}, J.~I., {et~al.} 2014, \natgeo,
  7, 493

\bibitem[{{Hofgartner} {et~al.}(2016){Hofgartner}, {Hayes}, {Lunine}, {Zebker},
  {Lorenz}, {Malaska}, {Mastrogiuseppe}, {Notarnicola}, \&
  {Soderblom}}]{hofgartner_etal_2016}
---. 2016, \icarus, 271, 338

\bibitem[{{Hosking} {et~al.}(1993){Hosking}, {Tonkin}, {Proykova}, {Hewitt},
  {McN Alford}, \& {Button}}]{hosking_etal_1993}
{Hosking}, M.~W., {Tonkin}, B.~A., {Proykova}, Y.~G., {et~al.} 1993, Supercond.
  Sci. Technol., 6, 549

\bibitem[{{Le Gall} {et~al.}(2016){Le Gall}, {Malaska}, {Lorenz}, {Janssen},
  {Tokano}, {Hayes}, {Mastrogiuseppe}, {Lunine}, {Veyssi{\`e}re}, {Encrenaz},
  \& {Karatekin}}]{legall_etal_2016}
{Le Gall}, A., {Malaska}, M.~J., {Lorenz}, R.~D., {et~al.} 2016, \jgr, 121, 233

\bibitem[{{Leifer} {et~al.}(2017){Leifer}, {Chernykh}, {Shakhova}, \&
  {Semiletov}}]{leifer_etal_2017}
{Leifer}, I., {Chernykh}, D., {Shakhova}, N., \& {Semiletov}, I. 2017,
  Cryosphere, 11, 1333

\bibitem[{{Leifer} {et~al.}(2015){Leifer}, {Solomon}, {von Deimling}, {Rehder},
  {Coffin}, \& {Linke}}]{leifer_etal_2015}
{Leifer}, I., {Solomon}, E., {von Deimling}, J.~S., {et~al.} 2015, Mar. Pet.
  Geol., 68, 806

\bibitem[{Lide(1974)}]{handbook74th}
Lide, D.~P., ed. 1974, CRC Handbook of Chemistry and Physics, 74th edn. (CRC
  PRESS)

\bibitem[{{Malaska} {et~al.}(2017{\natexlab{a}}){Malaska}, {Hodyss}, {Lunine},
  {Hayes}, {Hofgartner}, {Hollyday}, \& {Lorenz}}]{malaska_etal_2017}
{Malaska}, M.~J., {Hodyss}, R., {Lunine}, J.~I., {et~al.} 2017{\natexlab{a}},
  \icarus, 289, 94

\bibitem[{{Malaska} {et~al.}(2017{\natexlab{b}}){Malaska}, {Hodyss}, {Lunine},
  {Hayes}, {Hofgartner}, {Hollyday}, \&
  {Lorenz}}]{malaska_etal_2017_NASA_press_release}
---. 2017{\natexlab{b}}, Experiments Show Titan Lakes May Fizz with Nitrogen,
  \burl{https://www.nasa.gov/feature/jpl/experiments-show-titan-lakes-may-fizz-with-nitrogen},
  [Online; accessed 29-May-2017]

\bibitem[{{Mie}(1908)}]{mie_1908}
{Mie}, G. 1908, \AnnderPhys, 330, 377

\bibitem[{{Mitchell} {et~al.}(2015){Mitchell}, {Barmatz}, {Jamieson}, {Lorenz},
  \& {Lunine}}]{mitchell_etal_2015}
{Mitchell}, K.~L., {Barmatz}, M.~B., {Jamieson}, C.~S., {Lorenz}, R.~D., \&
  {Lunine}, J.~I. 2015, \georl, 42, 1340

\bibitem[{{Molina-Cuberos} {et~al.}(1999){Molina-Cuberos}, {L{\'o}pez-Moreno},
  {Rodrigo}, {Lara}, \& {O'Brien}}]{molina-cuberos_etal_1999}
{Molina-Cuberos}, G.~J., {L{\'o}pez-Moreno}, J.~J., {Rodrigo}, R., {Lara},
  L.~M., \& {O'Brien}, K. 1999, \pss, 47, 1347

\bibitem[{{Nougier}(1987)}]{nougier_1987}
{Nougier}, J.~P. 1987, {M\'{e}thodes de calcul num\'{e}rique} ({Paris}:
  {Masson})

\bibitem[{{Parrish} \& {Hiza}(1974)}]{parrish_hiza_1974}
{Parrish}, W.~R., \& {Hiza}, M.~J. 1974, Adv. Cryog. Eng., 19, 300

\bibitem[{{Poling} {et~al.}(2007){Poling}, {Prausnitz}, \&
  {O'Connell}}]{poling_2007}
{Poling}, B.~E., {Prausnitz}, J.~M., \& {O'Connell}, J. 2007, {The Properties
  of Gases and Liquids}, 5th edn. (Englewood Cliffs: McGraw-Hill Professional)

\bibitem[{{Prince} \& {Blanch}(1990)}]{prince_blanch_1990}
{Prince}, M.~J., \& {Blanch}, H.~W. 1990, \aichej, 36, 1485

\bibitem[{{S\'{a}nchez-Lavega}(2010)}]{sanchezlavega}
{S\'{a}nchez-Lavega}, A. 2010, An Introduction to Planetary Atmospheres (CRC
  Press)

\bibitem[{{Sprow} \& {Prausnitz}(1966)}]{sprow_prausnitz_1966}
{Sprow}, F.~B., \& {Prausnitz}, J.~M. 1966, AIChE J., 12, 780

\bibitem[{{Stephan} {et~al.}(2010){Stephan}, {Jaumann}, {Brown}, {Soderblom},
  {Soderblom}, {Barnes}, {Sotin}, {Griffith}, {Kirk}, {Baines}, {Buratti},
  {Clark}, {Lytle}, {Nelson}, \& {Nicholson}}]{stephan_etal_2010}
{Stephan}, K., {Jaumann}, R., {Brown}, R.~H., {et~al.} 2010, \georl

\bibitem[{{Stofan} {et~al.}(2007){Stofan}, {Elachi}, {Lunine}, {Lorenz},
  {Stiles}, {Mitchell}, {Ostro}, {Soderblom}, {Wood}, {Zebker}, {Wall},
  {Janssen}, {Kirk}, {Lopes}, {Paganelli}, {Radebaugh}, {Wye}, {Anderson},
  {Allison}, {Boehmer}, {Callahan}, {Encrenaz}, {Flamini}, {Francescetti},
  {Gim}, {Hamilton}, {Hensley}, {Johnson}, {Kelleher}, {Muhleman}, {Paillou},
  {Picardi}, {Posa}, {Roth}, {Seu}, {Shaffer}, {Vetrella}, \&
  {West}}]{stofan_etal_2007}
{Stofan}, E.~R., {Elachi}, C., {Lunine}, J.~I., {et~al.} 2007, \nature, 445, 61

\bibitem[{{Tan} {et~al.}(2013){Tan}, {Kargel}, \& {Marion}}]{tan_etal_2013}
{Tan}, S.~P., {Kargel}, J.~S., \& {Marion}, G.~M. 2013, \icarus, 222, 53

\bibitem[{{Vehkam\"{a}ki}(2006)}]{vehkamaki_2006}
{Vehkam\"{a}ki}, H. 2006, {Classical Nucleation Theory in Multicomponent
  Systems} (Berlin, Heidelberg: Springer)

\bibitem[{{Volmer} \& {Weber}(1926)}]{volmer_weber_1926}
{Volmer}, M., \& {Weber}, A. 1926, Zeit. Physik. Chemie, 119, 277

\bibitem[{{Wilke} \& {Chang}(1955)}]{wilke_chang_1955}
{Wilke}, C.~R., \& {Chang}, P. 1955, \aichej, 1, 264

\bibitem[{{Wye} {et~al.}(2009){Wye}, {Zebker}, \& {Lorenz}}]{wye_etal_2009}
{Wye}, L.~C., {Zebker}, H.~A., \& {Lorenz}, R.~D. 2009, \georl, 36, L16201

\bibitem[{{Zebker} {et~al.}(2014){Zebker}, {Hayes}, {Janssen}, {Le Gall},
  {Lorenz}, \& {Wye}}]{zebker_etal_2014}
{Zebker}, H., {Hayes}, A., {Janssen}, M., {et~al.} 2014, \georl, 41, 308

\bibitem[{{Zeldovich}(1943)}]{zeldovich_1943}
{Zeldovich}, J.~B. 1943, Acta Physicochimica, URSS, 18, 1

\end{thebibliography}

\begin{acknowledgements}
\noindent{\bf Acknowledgements}
Our understanding of the problem of bubble stream RADAR reflectivity has been greatly helped by discussions with
Dr. Jason Hofgartner of the Jet Propulsion Laboratory; it is our pleasure to acknowledge our indebtedness to him. We also thank 
the anonymous reviewer, who led us to significantly clarify  our manuscript.
\end{acknowledgements}

\appendix
\section{Model of Bubble Ascension and Growth}

   In a column of liquid, the gas bubbles have a vertical upward motion due to buoyancy forces, the liquid
flowing around bubbles rapidly reaches a high Reynolds number. In such a situation, the bubble velocity $U_b$ (m s$^{-1}$) 
can be estimated with \citep{clift_etal_1978}
\begin{equation}\label{Velocity_large_Re}
  U_{b} = \frac{2}{3} \sqrt{\frac{g a \Delta \rho}{\rho}}
\end{equation}
since, during their ascent to the free surface, the bubbles distort, the parameter $a$ represents a characteristic length
of bubble geometry. For the sake of simplicity, we adopted the approximation $a \simeq r_{b}$, with $r_{b}$ the bubble
``radius'' or typical size. In addition, we
have $\Delta \rho= \rho_{\rm liq} - \rho_{\rm gas} \simeq \rho_{\rm liq}$, then $\Delta \rho/\rho \sim 1$, leading to
\begin{equation}
  U_b \simeq \frac{2}{3} \sqrt{g_{\rm Titan} r_b}
\end{equation}
We emphasize that, before adopting the velocity given by Eq. (\ref{Velocity_large_Re}), we performed tests using the so-called
``Levich velocity''
\begin{equation}\label{levitchVelocity}
  U \simeq \frac{\rho \, g_{\rm Titan} \, r_b^2}{9\eta_{\rm liq}}
\end{equation}
where $\eta_{\rm liq}$ is the viscosity of the liquid, for which velocity is valid for relatively moderate Reynolds numbers, {\it i.e.} 
$50 \lesssim Re \lesssim 200$ \citep{clift_etal_1978}. In that case, the Reynolds numbers, obtained in our simulation,
quickly reached $\sim 10^3$, to finally increase to $\sim 10^5$ near the surface, far beyond the validity of Eq. (\ref{levitchVelocity}). 
We, then, turned to Eq. (\ref{Velocity_large_Re}) 
to get more consistent numerical simulations. The initial depth of $H_0 \sim 0.50$ m, found to get centimeter-sized bubbles at the surface, has to be
understood as a minimum. Indeed, in the early times of the ascent, the Reynolds numbers were below $\sim 200$ and Levich's form should have been employed during
this stage,
leading to larger $H_0$'s values.\\
   In the case of the fluid sphere, for high Reynolds numbers, the Sherwood number and the Peclet number 
are linked through the equation \citep{clift_etal_1978}
\begin{equation}
  \mathrm{Sh} = \frac{2}{\sqrt{\pi}} \mathrm{Pe}^{1/2}
\end{equation}
We recall that
\begin{equation}
  \mathrm{Sh} = \frac{k l}{D}
\end{equation}
where $k$ is the convective mass transfer rate (m s$^{-1}$), $l$ is a characteristic length (m), and $D$ is the molecular diffusion
coefficient (m$^2$ s$^{-1}$), in our context $D \sim D_{\rm N_2-CH_4}$. The Peclet number is given by
\begin{equation}
\mathrm{Pe} = \frac{U l}{D}
\end{equation}
Using the above equation and taking $l \sim r_b$, we can express that the convective mass transfer rate
are linked through the equation \citep{clift_etal_1978}
\begin{equation}
  k = \sqrt{\frac{2}{\pi}} \sqrt{\frac{D_{\rm N_2-CH_4} U_b}{r_b}}
\end{equation}
Here, the N$_2$ bubble content, noted as $n$ (mol) is driven by the equation
\begin{equation}
  \frac{\mathrm{d}n}{\mathrm{d}t} = k \, 4\pi r_b^2 \, \Delta c_{\rm N_2}
\end{equation}
This equation can be easily reformulated as
\begin{equation}\label{equa_A}
\frac{\mathrm{d}n}{\mathrm{d}h} = -\sqrt{\frac{2}{\pi}} \left(\frac{D_{\rm N_2-CH_4}}{r_b U_b}\right)^{1/2} 
           4\pi r_b^2 \, \Delta c_{\rm N_2}
\end{equation}
where $h$ (m) is the depth at which the bubble is located at a particular moment. For convenience, we have considered time as 
a function of $h$, which has been chosen as our independent variable. Thus, $t(h)$ follows the law
\begin{equation}\label{equa_B}
\frac{\mathrm{d}t}{\mathrm{d}h} = -\frac{1}{U_b}
\end{equation}
The external bubble pressure $P_e$ is ruled by the hydrostatic law $P_e(h)= P_0 + \rho g_{\rm Titan} h$ where
$P_0$ represents the atmospheric pressure at the sea surface, leading to
\begin{equation}\label{equa_C}
\frac{\mathrm{d}P_e}{\mathrm{d}h} = \rho g_{\rm Titan}
\end{equation}
With $P_i$ as the internal pressure of bubbles, assumed spherical, we can write the ideal gas law
\begin{equation}
  P_i(h) \times \frac{4}{3}\pi r_b^3 = n(h) R_{\rm gas} T
\end{equation}
with $R_{\rm gas}$ as the gas constant. The pressures $P_e$ and $P_i$ are linked by Laplace's equation
\begin{equation}
  P_i= P_e + \frac{2\sigma}{r_b}
\end{equation}
from this, we can easily derive the equation governing the evolution of the bubble radius
\begin{equation}\label{equa_D}
  \frac{\mathrm{d}r_b}{\mathrm{d}h} = -\frac{R_{\rm gas}T}{P_i} \sqrt{\frac{2}{\pi}} 
                   \left(\frac{D_{\rm N_2-CH_4}}{r_b U_b}\right)^{1/2} \Delta c_{\rm N_2} 
                          - \frac{\rho g_{\rm Titan} r_b}{3 P_i}
\end{equation}
In summary, we have four unknowns: $n(h)$, $t(h)$, $P_e(h)$, and $r_b(h)$, which are found by numerically integrating 
\citep{nougier_1987} the system of four equations: (\ref{equa_A})--(\ref{equa_C}) and (\ref{equa_D}).
Assuming an isothermal column of liquid, at temperature $T$, showing a uniform supersaturation $\Delta c_{\rm N_2}$ in
dissolved N$_2$, these equations are solved adopting a starting depth $H_0$ and an initial radius $R_0$ for bubbles.

\section{Diffusion coefficient of nitrogen in liquid methane}

The N$_2$ molecules, initially in the vicinity of a given microbubble, can migrate toward the bubble interior under the influence of thermal
agitation. The literature proposes several methods to estimate
the diffusion coefficient $D_{\rm N_2-CH_4}$ of the nitrogen molecule through liquid methane \citep{poling_2007}. Among these methods, 
the Wilke--Chang technique \citep{wilke_chang_1955} is widely used. It is based on correlations and provides diffusion
coefficient $D^{0}_{\rm A-B}$ of a compound A in a compound B, at infinite
dissolution, \textit{i.e.} when the mole fraction of A is very small. For our system, one can write
\begin{equation}\label{Dn2ch4_0}
 D^{0}_{\rm N_2-CH_4}= \frac{7.4 \times 10^{-11} (\Phi M_{\rm CH_4})^{1/2} T}{\eta_{\rm CH_4} V_{\rm N_2}^{0.6}}
\end{equation}
with $D^{0}_{\rm N_2-CH_4}$ in cm$^{2}$ s$^{-1}$, $\Phi$ is an adimensional coefficient around unity, $M_{\rm CH_4}$ is the molecular weight 
(g mol$^{-1}$) 
of methane, $\eta_{\rm CH_4}$ is the dynamic viscosity of liquid methane (Pa s), and $V_{\rm N_2}$ is the molar volume of solute N$_2$ at its 
normal boiling temperature (cm$^3$ mol$^{-1}$). The molecular weight has the well known value
$M_{\rm CH_4}= 16.04$ g mol$^{-1}$, the viscosity is provided by the literature \citep{hellemans_etal_1970} 
$\eta_{\rm CH_4} \simeq 1.7 \times 10^{-3}$ Pa s and the molar volume $V_{\rm N_2}= 35$ cm$^{3}$ mol$^{-1}$ \citep{handbook74th}.
At $T= 95$ K, these numbers lead to $D^{0}_{\rm N_2-CH_4}\simeq 2 \times 10^{-6}$ cm$^2$ s$^{-1}$. This determination is comparable
to those published for other simple molecules in the liquid state \citep{poling_2007}.\\
   Liquid methane, in equilibrium with a vapor dominated by nitrogen, such as in the case of Titan, should contain an amount of dissolved
nitrogen around $0.15$ in mole fraction (see Fig.~\ref{binaryN2CH4}). Then the assumption of infinite dissolution is not valid in our
context. Fortunately, empirical corrections are available and the diffusion coefficient $D_{\rm AB}$ can be derived from
coefficients $D^0_{\rm AB}$ and $D^0_{\rm BA}$ obtained in the frame of the hypothesis of infinite dissolution. For instance, one
may use \citep{poling_2007}
\begin{equation}\label{Dn2ch4_0}
 D_{\rm AB}= (D^0_{\rm BA} x_{\rm A} + D^0_{\rm AB} x_{\rm B}) \, \alpha
\end{equation}
where $x_{\rm A(B)}$ are the respective mole fraction and $\alpha$ is a thermodynamic coefficient, which is not too different from the unity.
Using an approach similar to the one previously done for nitrogen, we computed an estimation for the diffusion coefficient of methane in
liquid nitrogen, in the case of large dissolution, $D^0_{\rm CH_4-N_2} \simeq 4 \times 10^{-5}$ cm$^2$ s$^{-1}$, using
$\eta_{\rm N_2}\sim 10^{-4}$ Pa s \citep{forster_1963}. Our final estimation for the diffusion coefficient of N$_2$ in liquid CH$_4$
is $D_{\rm N_2-CH_4} \simeq 10^{-5}$ cm$^2$ s$^{-1}$.

\end{document}